# A Double Resonance Approach to Submillimeter/ Terahertz Remote Sensing at Atmospheric Pressure

Frank C. De Lucia, *Member IEEE*, Douglas T. Petkie, and Henry O. Everitt

*Abstract*–The remote sensing of gases in complex mixtures at atmospheric pressure is a challenging problem and much attention has been paid to it.  The most fundamental difference between this application and highly successful astrophysical and upper atmospheric remote sensing is the line width associated with atmospheric pressure broadening, ~ 5 GHz in all spectral regions.  In this paper, we discuss quantitatively a new approach that would use a short pulse infrared laser to modulate the submillimeter/terahertz (SMM/THz) spectral absorptions on the time scale of atmospheric relaxation.  We show that such a scheme has three important attributes:  (1) The time resolved pump makes it possible and efficient to separate signal from atmospheric and system clutter, thereby gaining as much as a factor of $10^6$ in sensitivity, (2) The 3-D information matrix (infrared pump laser frequency, SMM/THz probe frequency, and time resolved SMM/THz relaxation) can provide orders of magnitude greater specificity than a sensor that uses only one of these three dimensions, and (3) The congested and relatively weak spectra associated with large molecules can actually be an asset because the usually deleterious effect of their overlapping spectra can be used to increase signal strength.

## I. INTRODUCTION

There has been a long-standing interest in the remote spectroscopic detection and quantification of gases at or near atmospheric pressure.  While most of this activity has been in the much more technologically developed infrared, there has been considerable interest in the spectral region variously referred to as the microwave, millimeter wave, submillimeter wave, or terahertz [1-4], a region which we will refer to in this paper as the submillimeter/terahertz (SMM/THz).

Interestingly, the most highly developed and successfully fielded applications have also probed the most remote regions: studies of the interstellar medium [5] and the upper atmosphere [6].  However, there are important applications that involve sensing at, or near the ambient pressure of the earth's surface, and it is these applications that are addressed by this paper.  The principle physical characteristic that separates these terrestrial applications from the highly successful astronomical and upper atmospheric applications is the large linewidths associated with tropospheric pressure broadening.  At atmospheric pressure, these linewidths are ~ 5000 MHz, in comparison to their Doppler limited linewidths in this spectral region which are ~ 1 MHz [7].  These large linewidths have significant negative impact on both specificity and, although less generally recognized, on sensitivity as well.

The use of this spectral region and the molecular rotational fingerprint for point detection of gas is highly favorable because the pressure of the sample can be reduced so that the linewidths are reduced to the Doppler limit, ~ 1 MHz, well below the average spacing of spectral lines.  These narrow, well-resolved lines provide not only the redundant fingerprints of the molecular species, but also provide important means of separating the molecular signatures from other random and systematic system variations.  We have considered this application in some detail and will not further consider it here [8-10].

In this paper, we will consider the issues that make conventional remote sensing at tropospheric pressures *orders of magnitude more challenging* than the established astrophysical, upper atmosphere, and point gas sensing counterparts.  We will then quantitatively discuss a new approach that mitigates these challenges.

## II. THE CHALLENGES

In this section, we will introduce and briefly discuss the challenges that must be overcome for successful application of the SMM/THz spectral region to the terrestrial remote sensing of gases.

### A. *The Number of Available Channels*

The available spectral space for spectroscopic remote sensing is range dependent because of atmospheric transmission, but if one counts the number of ~ 5 GHz channels in the atmospheric windows, the number is fewer than 100.  As a result, unless the potential analytes in the atmospheric mixture include only a few light species that have

Manuscript received XXX, 2006.  This work was supported by the Army Research Office and the Defense Advanced Projects Research Agency.
  F. C. De Lucia is with the Dept. of Physics, Ohio State University, Columbus, OH 43210; phone 614-688-4774; fax 614-292-7557; fcd@mps.ohio-state.edu.
  D. T. Petkie is with the Department of Physics, Wright State University, Dayton, OH 45435.
  H. O. Everitt is with the Army Aviation and Missile Research, Development, and Engineering Center, Redstone Arsenal, AL 35898.

very sparse spectra, there are insufficient information channels for robust, molecule-specific detection.

B. *Species with Resolvable Lines or Bandheads*

While small molecules such as water and ammonia with very strong and sparse spectra are often used in THz demonstrations, most molecules of interest are heavier and have considerably more crowed spectra. In fact, at atmospheric pressure, molecules that are heavier than ~ 50 amu will have even prominent spectral structural features such as band heads washed out by the pressure broadened linewidths.

C. *Separation of Absorption Signal from Fluctuations*

The narrow linewidths associated with Doppler limited spectra provide a convenient signature for the rejection of fluctuations in time and/or frequency imposed by the sensor system. In SMM/THz spectrometers, interference effects notoriously cause frequency-dependent power variations of tens of percent even in well-designed systems. If there is no way to separate this effect from the molecular absorption, then the minimum detectable absorption will also be tens of percent.

Perhaps less broadly appreciated, temporal fluctuations of atmospheric attenuation, caused primarily by temporal and spatial variations of water concentration and temperature along the path, must also be separated from molecular absorption. While the time scale is highly scenario dependent, it is expected to be on the time scale of 1 s, which is comparable to a typical sensor integration time. Since ambient atmospheric SMM/THz attenuation is large, even in the transmission windows, this time-varying clutter can overwhelm weak spectra [11, 12].

D. *Comparison with the Point Sensor Approach*

To understand the impact of these effects, it is useful to compare this atmospheric situation with that of well established SMM spectrometers or point gas sensors based on low-pressure samples. Because electronic sources in this spectral region are inherently very quiet and there is little ambient black body radiation [13], very small fractional absorptions (typically 1 part in $10^7$ after 1 s of signal averaging) are routinely observed in Doppler-limited spectrometers [8, 9]. This is six orders of magnitude better than the limits discussed above for a SMM/THz absorption system operating in the real atmosphere. It is important at this point to remember that this difference is not due to any fundamental limits (which are scenario independent), but rather due to these *system and scenario* limitations.

## III. A NEW METHODOLOGY

To meet these challenges we discuss here a new methodology to provide: (1) More independent channels and/or kinds of information, and (2) A way to modulate the signal associated with the trace molecular species so that it can be separated from atmospheric clutter effects and noise.

Specifically, we propose an IR pump - SMM/THz probe double resonance technique in which a pulsed $CO_2$ TEA laser modulates molecular SMM/THz emission and absorption. Typical lasers produce pulses of duration 100 nsec or, if mode locked, pulses of duration ~100 psec. This modulation is then detected by a bore sighted THz transceiver.

The sensor could be configured to operate either in a general survey mode or optimized for a particular scenario. For the former, the laser would step thorough its available frequencies, i.e. the ~ 50 available $CO_2$ laser lines, while the probe monitored the SMM/THz response of the target molecular cloud. An attractive choice for the latter would be the ~10 independent (pressure broadened limited) channels in the 50 GHz wide atmospheric window centered at 240 GHz for which broadband frequency multiplier solid-state technology is commercially available. This probe could be stepped through the resolution channels. For the simultaneous observation of multiple probe channels, more elaborate multi-channel receiver technology similar to that used by the radio astronomy community is also a choice. For scenario specific implementations, this generality of pump and probe frequencies can be significantly reduced, but for the minimization of false alarms, particular attention would need to be paid to the signatures of potential clutter species.

In many ways, this is an atmospheric pressure version of an Optically Pumped Far Infrared (OPFIR) laser. Moreover, the TEA laser source is broad and tunable, and the target gas lines are broad as well because of the broadening due to the atmospheric pressure. Consequently, pump coincidence is much more easily achieved than for low pressure OPFIR lasers, and the proposed technique should be quite general for any molecule with IR absorption in the 9-11 micron region.

At a fundamental level, the technique depends on the rapid molecular collisional relaxation time, which at atmospheric pressure and temperature is ~ 100 psec and only weakly depends on molecular mass and dipole moment. Thus, the short 100 psec laser pulses and rapid collisional relaxation modulate the molecular THz emission or absorption on a time scale much faster than the 1 s temporal atmospheric fluctuations, making it straightforward to separate the molecular signal from signals due to clutter. Furthermore, the time-modulated spectra exhibit a molecule-unique pattern of enhanced and reduced absorption that can be resolved from the atmospheric baseline, as will be discussed below.

A. *A specific spectroscopic example $^{13}CH_3F$*

Consider as an example the application of this technique to trace amounts of $^{13}CH_3F$ in the atmosphere. Because $^{13}CH_3F$ is a well known OPFIR laser medium, it is well studied and the parameters for a quantitative analysis are available [14-16]. However, we will show below that the proposed methodology is not particularly dependent upon its favorable OPFIR laser properties. Fig. 1 shows both the energy levels involved and the impact of the pump laser on the spectral signature. The 9P(32) line of the TEA $CO_2$ laser excites the R-branch ($\Delta J = +1$) transition from the heavily populated $J = 4$ level of the ground (v = 0) vibrational state of $^{13}CH_3F$ to the nearly empty $J = 5$ level of the first excited (C-F stretch) vibrational state (v = 1). Specific rotational transitions in both the ground and excited vibrational state are population modulated by the TEA laser pump, which can be observed as modulated SMM/THz molecular emission and absorption signals.

Specifically, the top right trace of Fig. **1** shows that excitation by a TEA $CO_2$ laser can result in enhanced absorption or emission on the $J = 4 - 5$ transitions near 0.25 THz for molecules in both the $v = 0$ and 1 vibrational states. For each $J = 4 - 5$ transition, the large pump and atmospheric (~ 5 GHz) linewidths cause *all five* of the $K$ states (the spacing from $K = 0$ to $K = 4$ is 68 MHz) to be pumped simultaneously, adding to the strength of the SMM/THz signal. Furthermore, because the $J = 4 - 5$ frequencies in one vibrational state differ from those in the other vibrational state by ~ 3 GHz, less than the atmospherically pressure-broadened linewidth, these two composite THz signals overlap and add as well. Most importantly, the analyte signatures can easily be separated from atmospheric clutter because the millisecond or longer atmospheric fluctuations are effectively frozen on the nanosecond analyte modulation time scale.

Molecule specificity arises because the molecular fingerprint requires both a coincidence between the laser pump wavelength and IR molecular absorption *and* a modulated THz signal at specific frequencies: In this case the emission at 250 GHz and the enhanced absorptions due to the $J = 5 - 6$ $v = 1$ transition near 300 GHz and the $J = 3 - 4$ $v = 0$ transition near 200 GHz. This tripartite signature will last for the relaxation time of the atmosphere (100 psec), which may also contain a temporal signature. In the case of methyl fluoride, each feature will be easily resolved because the spectral width of ~5 GHz is smaller than the $2B = 50$ GHz spacing of the features.

Now consider a much heavier species, with smaller rotational constants. While the rotational selection rules and spectroscopy of molecules, especially asymmetric rotors, is both complex and molecule specific, some general observations can be made. Because each asymmetric rotor has limiting prolate and oblate symmetric top bases, the general character (the density, strength, and general location in frequency) of its spectrum can be established by consideration of their symmetric top limits. Because in the transition from the symmetric top base to the asymmetric top base there is a generalization that the direction of the dipole moment can lie along any of the principal axes of intertia, the transition frequencies of strong lines are separated by ~$2R$ (where $R$ is *any* one of the rotational constants $A$, $B$, or $C$).

If, as in the case of $CH_3F$, the appropriate $R$ is $B$ - with $2B =$ 50 GHz (considerably larger than the atmospheric pressure broadening width of ~ 5 GHz), the full tripartite modulation shown in Fig. 1 is realized. However, as the offset between the emission and absorption features of the signature approaches the pressure broadened linewidth, there is an overlap and the net modulation efficiency is reduced as shown in Fig. 2. Figure 3 shows this quantitatively. Because molecular size affects both vapor pressure and the size of rotational constants, there is a correlation between molecules that have vapor pressure and those for which the modulation efficiency is reasonably favorable. To within wide spectroscopic and vapor pressure variability, this gas phase limit might occur for the case for which the emission/absorption off set ~ 2.5 GHz ~ one half of the pressure broadened linewidth and for which the modulation efficiency is only reduced by a factor of ~5.

Additionally, we will show in Section IV.A that as the rotational constants are reduced, molecular absorptions averaged over atmospheric pressure broadening linewidths will actually increase, thereby at least partially compensating for possible loss in modulation efficiency.

B. *Requirements for excitation laser:* To assess the feasibility of this technique, let us consider quantitatively some of the required $CO_2$ TEA laser characteristics. In terms of energy these range from relatively small laboratory systems of energy ~0.1 J/macropulse to more complex, but 'compact' systems which produce ~100 J/macropulse [17]. The pulse structure of these systems also has a wide range; including the native macropulse duration of ~ 100 ns, the generation of multigigawatt pulse trains with micropulse widths of the order of 1 ns [18], passively mode-locked micropulses of duration ~150 psec [19], the production of terawatt micropulses of duration 160 psec [17], and the amplification and generation of micropulses whose duration was less than 1 psec [20].

Efficient pulse train production methods approximately conserve energy [18, 21]. Thus, for the purposes of this discussion we will assume that an appropriate modelock, either in a single oscillator or in a master oscillator - slave configuration, will be used to convert a 100 ns macropulse into a train of 10 micropulses, each of 100 psec duration and separated by 10 ns. For this pulse sequence, the peak power of each micropulse will range from $10^8$ W for the small 0.1 J laboratory system to $10^{11}$ W for the 100 J system.

The pump intensity must be high enough that the Rabi frequency is comparable to the atmospheric relaxation rate so that significant population transfer can take place from the rotational level in the ground vibrational state to the pumped rotational level in the upper vibrational state. In MKS units, the Rabi frequency is [22]

$$\omega_R = \frac{\mu E}{\hbar} = 8.75 \times 10^4 \sqrt{I[W/m^2]}, \quad (1)$$

where $E$ is the electric field and $I$ the laser intensity in units of Watts/m$^2$. Assume an IR molecular transition whose dipole moment $\mu = 0.1$ D, typical for $^{13}CH_3F$ and other molecules. Then if $10^9$ W from a 1 Joule system is spread over 10 cm$^2$, $\omega_R$ ~ 87 GHz, and a π excitation pulse $\tau_\pi = \pi / \omega_R$ would last ~ 35 psec. If the power from the 100 J system were spread over a 1 m$^2$ cross section, the corresponding Rabi frequency and π-pulse length would be 28 GHz and 110 psec, respectively.

To obtain the collisional relaxation rate, we can assume that the dominant atmospheric collision partners for the trace gas in question are nitrogen and oxygen. For these gases, the rotational relaxation rates will be comparable to the pressure broadening rate of ~ 3 MHz/Torr HWHM. This corresponds to a linewidth of $\Delta \nu = 5$ GHz and a mean time between collisions ($\tau_c = 1/\pi\Delta\nu$) of 60 psec. Thus, the mean collision time is comparable to the π pulse lengths estimated above.

C. *Sensitivity*: In the microwave (and SMM/THz as well) spectral region under equilibrium conditions there is a

significant population not only in the lower state of the transition, but in the upper state as well. Accordingly, the usual calculation of molecular absorption strength includes not only the absorption of molecules that are promoted from the lower state of the transition to the upper state, but also the largely canceling effect of the emission from molecules that make a transition from the upper state to the lower state. Quantitatively, in the usual equilibrium absorption coefficient, this emission from the upper state cancels all but (1-exp (-$h\nu/kT$)) ≈ $h\nu/kT$ of the absorption from the lower state [7]. However, in our non-equilibrium double resonance case, the π-pulse pump only places population in one state and the emissions/absorptions are stronger by the inverse of this factor - about 20 at 300 GHz.

In a pure sample, the absorption coefficient for one of the K components of the J = 4 - 5 transition of the $CH_3F$ rotational transitions near 300 GHz is ~ $10^{-2}$ cm$^{-1}$[7]. Because the pressure broadening of $CH_3F$ in oxygen or nitrogen is only about 1/5 of its self broadening (15 MHz/Torr), its peak absorption coefficient in the atmosphere is about 5 times larger. Thus, a sample dilution of $10^{-6}$ (1 ppm) over a 100 m ($10^4$ cm) path would yield a *modulated* signal absorption

$$\alpha_{pumped} l = \frac{\frac{\Delta\nu_{CH_3F}}{\Delta\nu_{air}} \times \alpha_{CH_3F} \times path \times \frac{kT}{h\nu}}{dilution}$$
$$= \frac{(5)(10^{-2} cm^{-1})(10^4 cm)(20)}{10^6} \quad (2)$$
$$= 10^{-2}$$

a very large signal relative to the noise sensitivity of a THz probe system. However, it is not large in comparison to the atmospheric clutter variations or to the 'absolute calibration' of the THz probe system. Thus, it is enormously more advantageous to lock onto this modulation signature than to try to deconvolute this absorption from the very large absorptions $\alpha l \sim 1$ (primarily due to water vapor) which are fluctuating in the atmosphere on a time scale of 1 s.

*This is one of the three main points of this paper.*

D. *Uniqueness of Signature*

One of the problems with infrared (IR) based remote detection systems is that the molecular absorption contours, which arise from the unresolved rotational structure, aren't very molecule specific. The same would be true for an ordinary THz sensing system because both would have the same pressure broadened linewidth of ~ 5 GHz. In fact, a widely tunable high-resolution infrared system (e. g. a *large* FTFIR) would have the advantage of looking at several different vibrational bands. However, the methodology that we present here provides a 3-D signature space that is much more molecule specific. The axes of this signature space are:
  a. The frequency of the THz signature.
  b. The frequency of the IR pump.
  c. The time of the relaxation between a. and b.

An example of a 3-D specificity matrix is shown in Fig. 4. The probe axis is the SMM/THz frequency. Its number of SMM/THz resolution elements will be limited to the number of 5 GHz wide channels that can be accommodated by whatever atmospheric transmission windows are available at the range of the scenario. This would be the number of channels available in a more conventional SMM/THz remote sensing system, increased by a factor that is determined by the gain attributable to the modulation against clutter and the $kT/h\nu$ ~20 enhancement that result from the action of the infrared pump.

The pump axis would have the same number of resolution elements as a similar purely infrared remote sensing system. This number would be determined by the overlap between the pump and the infrared vibrational-rotation transitions. For both the infrared and the double resonance system described here, the tunability of the TEA laser is therefore important.

The third axis, the temporal relaxation is more difficult to characterize because it represents an experimental regime that has not been explored; the rotation-vibrational relaxation of the analytes in air at high pressure. The closest work on this subject has been on low pressure gases in the context of optically pumped infrared lasers [16]. The primary decay observed by the probe will be most closely related to a state's total depopulation rate, which in turn is a sum over many constituent rates. Therefore, we expect that there will be some signature associated with this axis but that it will not represent as many independent points as the frequency axis. Typically, a simple exponential decay will be observed, whose characteristic time can be measured to ±10% but may only vary over about a factor of two from state to state. Weaker signatures may be considerably more specific.

*The much larger number of signature points in this 3-D matrix, as opposed to traditional spectroscopy which considers only the number on one of its 1-D axes, is the second of the three main points of this paper.*

E. *Remote sensing*

In remote sensing applications, the performance of the system described here will be a complex function of system parameters and scenario. Here we will discuss two specific examples, based on the 250 GHz $CH_3F$ signature shown in Fig. 1, to provide baselines, as well as some of the derivatives from these baselines to provide a broader quantitative sense.

In order to explore the edge of this performance, we will consider two limits of a challenging scenario, the detection with good specificity of a 100 m cloud 1 km away, with a gas dilution of 1 ppm. One km is probably near the maximum range imposed by atmospheric SMM/THz propagation at 250 GHz.

We will assume a SMM/THz probe which provides a diffraction limited ~ 1 m diameter beam at 1 km with a 1 meter antenna. This beam width corresponds to the example of the 100 J laser discussed in section III.B. In that example, when the TEA laser beam was expanded to fill a 1 m cross section, the resultant Rabi frequency was longer than the relaxation time by about a factor of two, which would reduce the pump efficiency also by about a factor of two. Alternatively, one could reduce the pump beam diameter relative to the probe beam to make the π-pulse lengths and the atmospheric relaxation times match, but at a similar cost in filling factor.

We also need to make a SMM/THz scenario assumption. Path geometries might include a background retro-reflector, direct transmission through the volume of interest, or a background with diffusive scattering.

For the purposes of baseline discussions, we will assume a cw 1 mW SMM/THz active illuminator for the retro-reflector/direct transmission scenario and a 60 W pulsed extended interaction oscillator (EIO) for the diffuse scatter scenario. More than 1 mW can currently be produced by solid-state sources, and significant development programs are currently underway that have targets for compact sources at least two orders of magnitude higher. The EIO is similar to those used for radar experiments in the SMM/THz, which in many cases also have to depend upon backscatter for their signals.

1. For the most optimistic direct retro-reflector geometry, if we assume a 10 db two-way atmospheric propagation loss at 250 GHz, the signal returned to the receiver will include a carrier of ~$10^{-4}$ W overlaid by whatever modulation arises from the molecular interaction. The 1% absorption calculated in Eq. 2 needs to be modified by a factor of two, because of the two way path, and by another factor of five because the K = 0, 1, 2, 3, and 4 components of the J = 4 - 5 transition will be degenerate at atmospheric pressure. Together these provide a 10% modulation and an absorbed power of $10^{-5}$ W. With the factor of two that resulted from the under pumping/filling factor effect from above, this reduces the absorbed power to ~5 x $10^{-6}$ W.

A heterodyne receiver has a noise power of $P_N = kT_N(Bb)^{1/2}$ where $T_N$ is the noise temperature of the SMM/THz receiver, $b$ the IF bandwidth, and $B$, the post detection bandwidth. For $T_N$ = 3000 K and $b = B = 10^{10}$ Hz (the bandwidth required for the mode-locked case), $P_N$ = 5 x $10^{-10}$ W.

However, since we are seeking to detect a small change in power in the returned signal (the carrier), we must also consider the noise $P'_n$ associated with the mixing of the blackbody noise with the carrier of the returned signal [24]. For this case

$$P'_n \approx \sqrt{kT\Delta\nu P_c} = \sqrt{(5\times10^{-11})(10^{-4})} \approx 7\times10^{-8} W \qquad (3)$$

As above, the absorption provides a signal level of 5 x $10^{-6}$ W, about 70 times $P'_n$. However, in the mode-lock process, each macropulse of the TEA laser, produces ~10 micropulses, so with a 10 Hz TEA laser, in one second of integration we have a S/N of 700.

2. At 1 km, the diffuse backscatter limit is less favorable, but of considerable importance because of the wider range of scenarios that it can encompass. For diffusive backscattering at one kilometer, only $10^{-6}$ of the SMM/THz probe will be backscattered into the 1 m receiver antenna, and with a 10% reflection efficiency, the probe signal is further reduce to ~$10^{-7}$ relative to the direct transmission and retro-reflector geometries. Even with the EIO source, the received carrier is reduced to 6 x $10^{-7}$ W, and the absorbed power is now 6 x $10^{-8}$ W. However, the mixing noise of Eq. 3 is also reduced, in this example to ~5 x $10^{-9}$ W. This provides for a single 100 ps pulse a S/N ~ 12, or for one second of integration a S/N ~120.

F. *The trade space*

A general purpose implementation which can probe a number of SMM/THz frequencies simultaneously so as to take full advantage of the specificity matrix shown in Fig. 2 would be complex, with the required array of EIOs - or would await the success of one of many THz technology initiatives. Accordingly, we would like briefly to discuss the trade spaces involved to understand this double resonance approach in the context of current and projected SMM/THz technology.

1. If rather than a S/N of 120 in the diffuse scattering scenario, a S/N of 2 is acceptable, considerably less SMM/THz probe power is needed. Since this reduction in probe carrier power takes us near to the receiver noise limit rather than the mixing noise limit, we can do the simpler calculation based on the receiver noise of 5 x $10^{-10}$ W. Then the $10^{-9}$ W absorption required at the receiver for the S/N of 2 must be increased by $10^7$ to account for the scattering loss, $10^1$ to account for the atmospheric loss, and 10 to account for the 10% absorption. This results in a required SMM/THz probe power of 1 W. For 1 second of integration time, this is reduced to 0.1 W.

2. In many scenarios, more than 1 second is available to make the observation. In receiver noise limited scenarios, 100 seconds would reduce the required SMM/THz power by 10; in mixing noise limited scenarios by a factor of 100.

3. At 100 m, for diffuse back scattering scenarios, the required SMM/THz power is reduced by a factor of 100, because the return signal from diffuse backscattering decreases with the square of the distance.

4. Finally, there may be objects in the background that provide signal returns larger than that of a diffusive reflection. Indeed, it is this effect that results in the well known large dynamic range in active images that combine both 'glint' and diffuse returns.

Collectively, these provide a large and attractive design space. While the last of these items is not quantifiable, he first three taken together reduce the required SMM/THz power by ~$10^5$ - $10^6$ from the 60 W EIO limit to levels well below 1 mW. Any SMM/THz power available above this could be traded for shorter integration, higher S/N, longer range, and unaccounted for system losses.

## IV. HOW SPECIAL IS CH$_3$F? WHAT ABOUT LARGER MOLECULES?

Clearly CH$_3$F is not a large molecule. What happens as we go toward many of the species of interest? In this section, we will discuss the third important point of this paper:

*Because of the spectral overlap at high pressure, many of the rotational partition function problems associated with large molecules are either reduced or in some cases even turned into advantages.*

A. *Rotational Considerations*
For simplicity, we will consider symmetric-top molecules. While most of the species of interest are asymmetric rotors,

their spectral density and intensity are similar to those of symmetric tops. The main difference - the complexity of the high-resolution asymmetric rotor spectrum - is not important at the low resolution of atmospheric spectra.

Although one might expect the larger rotational partition function (which reduces the strength of each *individual* spectral line) to degrade the proposed scheme for larger molecules significantly, it doesn't because at a given frequency the increased partition function is largely cancelled because more $K$ levels exist within the pressure broadened linewidth. This is because there are $2J+1$ $K$ levels for each $J$ and, on average, the value of $J$ in a spectrum at a chosen frequency is *inversely* proportional to the rotational constants. Townes and Schawlow [24] calculate this summed coefficient to be

$$\alpha_{\text{total}} = \frac{2\pi h^2 N f_v}{9c(kT)^2 B} \sqrt{\frac{\pi Ch}{kT}} \mu^2 \frac{(4J+3)(J+2)}{(J+1)^2} \frac{v_o^2 \Delta v}{(v-v_o)^2+(\Delta v)^2} \quad (4)$$

where $N$ is the number density of the molecules, $h$ is Planck's constant, $f_v$ is the fraction in the vibrational state of interest (see the next section), $B$ and $C$ are rotational constants inversely proportional to the moments of inertia, $\mu$ the dipole matrix element, $\Delta v$ the linewidth, $v_0$ the transition frequency, and $T$ the temperature.

This relation shows that at worst the overlapped pressure broadened absorption is a slow function of frequency and actually would appear to increase for larger molecules with smaller rotational constants.

B. *Vibrational Considerations*

This is more difficult to evaluate in general and will depend on the details of the particular species.

a. For this scheme to work, there must be a vibrational band within the *tunable* range (~9-11 μm) of the TEA laser. For large molecules with many vibrational modes, it is highly probable that there will be vibrational bands in this range. These bands may be weaker or stronger absorbers (more will be weaker than stronger, but to some extent it will be possible to choose among several) and the required pump power will scale accordingly.

b. Especially in large molecules, there may be many low-lying torsional or bending modes. If some number of these lie at or below $kT$, $f_v$ will be reduced and absorptions such as calculated in Eq. (4) will be reduced accordingly. However, because of the pressure broadening (which is probably greater than the vibrational changes in the rotational spectra) this won't matter too much. Molecules will be pumped out of whatever low-lying mode they are in and promoted ~1000 cm$^{-1}$ to a corresponding combination band. All of these smaller absorptions will overlap and the result described by Eq. (4), with $f_v = 1$, will be restored by the sum.

V. DISCUSSION

Above we have discussed quantitatively a proposed methodology for remote sensing of trace gases at ambient atmospheric pressure. Here we discuss some of its scientific and technological unknowns in the context of its overall prognosis.

A. While the underlying double resonance physics of systems at low pressures typical of optically pumped lasers have been extensively studied [14-16, 25], atmospheric pressure is 4 to 5 orders of magnitude higher. Other than accelerated collisional processes and correspondingly fast pump/probe schemes, there are no first order modifications of the low-pressure behavior that would result from this higher pressure.

B. As an example of a potential obfuscating effect, non-linearities or other higher order effects might cause the TEA laser to modulate the IR and SMM/THz transmission of the atmosphere's ambient constituents. For example, although the water absorption bands in the atmosphere are far removed in frequency from the $CO_2$ pump, water is abundant and the bands are strong.

C. Many of the molecules of interest are large (> 100 amu). Because these species were inappropriate for low pressure OPFIRs, their spectroscopy and collision dynamics have been little studied. However, as noted above, they may be relatively much more advantageous for this atmospheric pressure application than in ordinary OPFIR lasers. While the calculations above appear favorable, this is an unknown spectroscopic frontier that would be worthy of early study in any pursuit of this approach.

D. The proposed double resonance scheme may also be useful with an infrared probe. The fundamental power of the proposed scheme lies primarily in the use of the TEA laser pump to modulate the atmosphere on a time scale related to its relaxation. The advantageous role of spectral overlap discussed in section IV is independent of the probe. At long range, an infrared probe would allow a much smaller probe beam diameter and significantly reduce the TEA laser power requirements. On the other hand, the SMM/THz is quieter, and it is possible to build room temperature receivers whose sensitivity are within an order of magnitude of even the limits set by these low noise levels.

VI. CONCLUSIONS

The use of the rotational signature of molecules for remote sensing applications is a problem that has received considerable attention and thought. In this paper, we have considered the impact of the fundamental difference, pressure broadening, between this atmospheric pressure application and highly successful low-pressure applications in astrophysics, atmospheric science, and the laboratory. We have used this analysis to put forth a double resonance methodology with three important attributes relative to other proposed and implemented techniques:

(1) The time resolved pump makes it efficient to separate signal from atmospheric and system clutter, thereby gaining a very large factor in sensitivity. This is very important and not often discussed.

(2) The 3-D information matrix (pump laser frequency, probe frequency, and time resolved molecular relaxation) can provide orders of magnitude greater specificity than a sensor that uses only one of these three.

(3) The congested and relatively weak spectra associated with large molecules can actually be a positive because the usually negative impact of overlapping spectra can be used to increase signal strength.

**Frank C. De Lucia** is a University Professor and Professor of Physics at Ohio State University and previously was Professor of Physics at Duke University. He has served both departments as Chairman. Along with his students and coworkers he has developed many of the basic technologies and systems approaches for the SMM/THz and exploited them for scientific studies. Among his research interests are imaging and phenomenology, remote sensing, the spectroscopy of small, fundamental molecules, SMM/THz techniques, collisional processes and mechanisms, the excitation and study of excited states, molecules of atmospheric and astronomical importance, and analytical chemistry and gas sensing. He is a member of the Editorial Board of The Journal of Molecular Spectroscopy; belongs to the American Physical Society, the Optical Society of America, the Institute of Electronic and Electrical Engineers, and Phi Beta Kappa. He was awarded the 1992 Max Planck Research Prize in Physics and the 2001 William F. Meggers Award of the Optical Society of America.

**Douglas T. Petkie** received a B.S. in physics from Carnegie Mellon University, Pittsburgh, PA in 1990 and a Ph.D. in physics from Ohio State University, Columbus, OH in 1996. He is currently an assistant professor of physics at Wright State University, Dayton, OH. His research interests include the development of submillimeter and terahertz systems for in-situ and remote sensing applications that utilize spectroscopy, imaging and radar techniques. He is a member of the American Physical Society, American Association of Physics Teachers, Council on Undergraduate Research, and Sigma Pi Sigma.

**Henry O. Everitt** is an Army senior research scientist at the Aviation and Missile Research, Development, and Engineering Center located at Redstone Arsenal, AL. He is also an Adjunct Professor of Physics at Duke University and the University of Alabama, Huntsville. His early work focused on the development and understanding of optically pumped far infrared lasers through the use of time-resolved THz/IR pump-probe double resonance techniques. More recently, he has concentrated on ultrafast optical studies of relaxation dynamics in wide bandgap semiconductor heterostructures and nanostructures. He is a member of the Editorial Board for the journal Quantum Information Processing and belongs to the American Physical Society, the Optical Society of America, the American Academy for the Advancement of Science, Sigma Xi, and Phi Beta Kappa. In 2004 he became a Fellow of the Optical Society of America and the Army Research Laboratory (Emeritus).

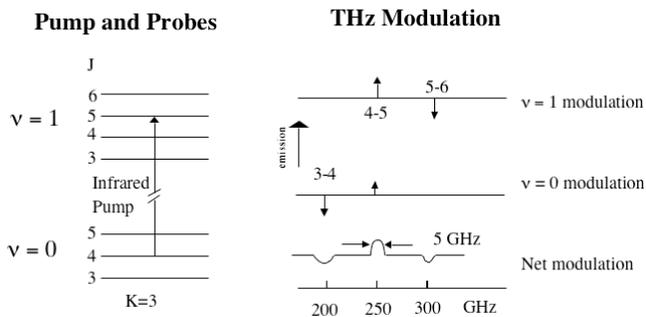

Fig. 1. The energy level diagram at the left shows that the pump connects J = 4 in the v = 0 state with J = 5 in the v = 1 state. The figure on the right shows the effect of the pump on the SMM/THz probe.

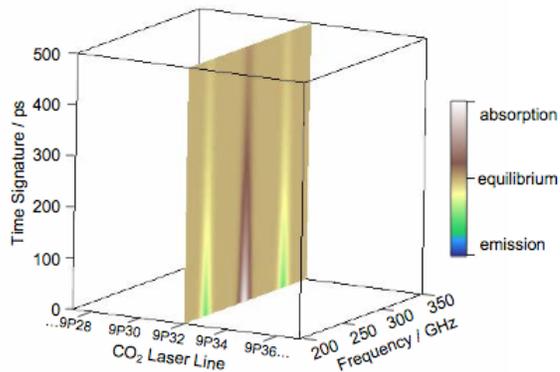

Fig. 4. The 3 - D specificity matrix associated with the time resolved double resonance scheme for the $^{13}CH_3F$ example. This includes only the signature that is known from low pressure OPFIR studies. At atmospheric pressure, other pump coincidence may be found. If so, this would add additional planes to this figure, one for each additional pump coincidence. Only the first order relaxation signatures are shown. There will be many other weaker signatures associated with the complexities of rotational relaxation.

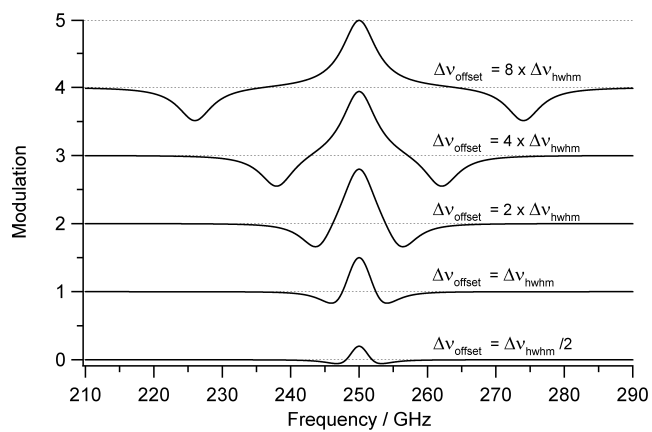

Fig. 2. SMM/THz signature as a function of separation of the pump induced absorption and pump induced emission in units of pressure broadened line width.

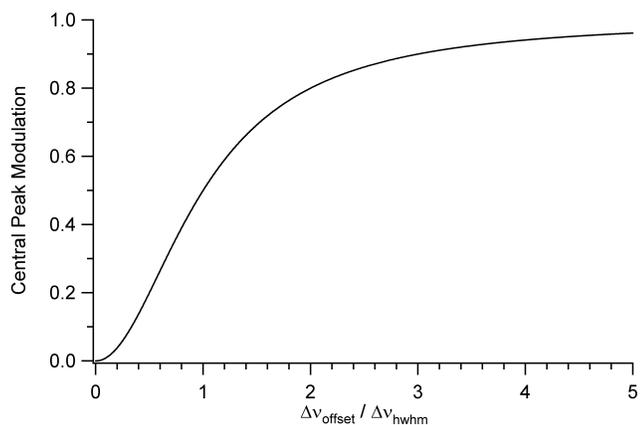

Fig. 3. The reduction in net modulation amplitude as a function of ratio of the signature offset to the pressure broadened linewidth.

8